%Paper: hep-ph/9306316
%From: JIZHI@PHYS.TAMU.EDU
%Date: Fri, 25 Jun 1993 12:30:49 CDT
%Date (revised): Fri, 25 Jun 1993 15:58:30 CDT

\magnification 1200
\baselineskip=12pt
\overfullrule=0pt
\nopagenumbers
\hsize=6.25truein
\vsize=9.5truein

\input tables.tex
\def\dg{\hbox{$^\dagger$}}

\def\lam{\hbox{$\lambda\kern-6pt^{\_\_}$}}

\def\r{\vbox{\hbox{\raise.5mm\hbox{$>$}}
\kern-18pt\hbox{\lower1.5mm\hbox{$\sim$}}}}
\def\l{\vbox{\hbox{\raise.5mm\hbox{$<$}}
\kern-18pt\hbox{\lower1.5mm\hbox{$\sim$}}}}
\def\one{\phantom{0}}

\def\at{\hbox{$^{\ast}$}}
\def\leaderfill{\leaders\hbox to 1em{\hss.\hss}\hfil}
\def\sqr#1#2{{\vcenter{\vbox{\hrule height.#2pt
        \hbox{\vrule width.#2pt height#1pt \kern#1pt
          \vrule width.#2pt}
        \hrule height.#2pt}}}}

\line{\hfil CTP-TAMU-23/93}
\line{\hfil NUB-TH-3062/92}
\line{\hfil SSCL-Preprint-229}
\vskip .50truein

\centerline{SUPERGRAVITY GRAND UNIFICATION, PROTON}
\centerline{DECAY AND COSMOLOGICAL CONSTRAINTS}
\bigskip

\centerline{R. Arnowitt{\footnote\dg{Talk at Les Rencontres de Physique de la
Vallee D'Aoste}}$^{a,b}$ and Pran Nath$^c$}
\centerline{{\footnote\at{PermanentAddress}}$^a$Center for Theoretical Physics,
Department
of Physics}
\centerline{Texas A\&M University, College Station, TX  77843-4242}
\centerline{$^b$Physics Research Division, Superconducting Super Collider}
\centerline{Laboratory, Dallas, TX  75237}
\centerline{$^c$Department of Physics, Northeastern University,}
\centerline{Boston, MA  02115}
\vskip 1.5truein

\centerline{ABSTRACT}
\bigskip

Properties and experimental predictions of a broad class of supergravity grand
unified
models possessing an $SU(5)$-type proton decay and $R$ parity are described.
Models of
this type can be described in terms of four parameters at the Gut scale in
addition to
those of the Standard Model i.e. $m_o$ (universal scalar mass), $m_{1/2}$
(universal
gaugino mass), $A_o$ (cubic soft breaking parameter) and $\tan \beta =
<H_2>/<H_1>$.  Thus
the 32 SUSY masses can be expressed in terms of $m_o, m_{1/2}, A_o \tan \beta$
and the as
yet unknown t-quark mass $m_t$.  Gut thresholds are examined and a simple model
leads to
grand unification consistent with $p$-decay data when $0.114 < \alpha_3 (M_z) <
0.135$, in
agreement with current values of $\alpha_3 (M_Z)$.  Proton decay is examined
for the
superheavy Higgs triplet mass $M_{H_3} < 10 M_G (M_G \simeq 1.5 \times
10^{16}$~GeV) and
squarks and gluinos lighter than 1 TeV.  Throughout most of the parameter space
chargino-neutralino scaling relations are predicted to hold:  $2
m_{\tilde{Z}_1} \cong
m_{\tilde{W}_1} \cong m_{\tilde{Z}_2}, m_{\tilde{W}_1} \simeq (1/4)
m_{\tilde{g}}$ (for
$\mu > 0$) or $m_{\tilde{W}_1} \simeq (1/3) m_{\tilde{g}}$ (for $\mu < 0$),
while
$m_{\tilde{W}_2} \cong m_{\tilde{Z}_3} \cong m_{\tilde{Z}_4} >>
m_{\tilde{Z}_1}$.  Future
proton decay experiments combined with LEP2 lead to further predictions, e.g.
for the
entire parameter space either proton decay should be seen at these or the
$\tilde{W}_1$
seen at LEP2.  Relic density constraints on the $\tilde{Z}_1$ further constrain
the
parameter space e.g. so that $m_t < 165$~GeV, $m_h < 105$~GeV, $m_{\tilde{W}_1}
<
100$~GeV and $m_{\tilde{Z}_1} < 50$~GeV when $M_{H_3}/M_G < 6$.
\vfill\eject

\baselineskip=18pt
\noindent
1.  INTRODUCTION
\medskip

Over the past two years, there has been considerable effort to deduce
consequences of
supergravity grand unification models$^{1)-7)}$.  This activity has been
stimulated in part
by the observation by several groups$^{8)}$ that unification of the coupling
constants
$\alpha_1 \equiv (5/3) \alpha_Y$, $\alpha_2$~and~$\alpha_3$ appears to occur at
a common
value $\alpha_G \simeq 0.04$ at a scale $M_G \approx 10^{16}$ GeV if one
assumes that the
particle spectrum below $M_G$ is the minimal supersymmetric one with just two
Higgs
doublets with the SUSY particles in the mass range $M_S \approx 10^{2-3}$ GeV.
Thus, while
unification fails by over 7 std. for the Standard Model mass spectrum, the SUSY
mass
spectrum introduces additional thresholds which allows grand unification to
occur.

A second impetus to the study of supergravity models is the possibility of
testing them
experimentally at current or future experiments.  The reason for this is due to
two
remarkable features of these models.  First, supergravity unification allows
for
spontaneous breaking of supersymmetry in the ``hidden'' sector, something that
is difficult
to achieve satisfactorily in low energy global supersymmetry, and remains an
important
unresolved problem in superstring theory.  While the physics of the hidden
sector is
unknown, it turns out that it can be characterized by just a few ``soft
breaking''
parameters$^{9,10)}$.  The second important feature is that the spontaneous
symmetry
breaking of supersymmetry can then trigger the breaking of $SU(2) \times
U(1)^{11)}$.  The
most theoretically appealing way of doing this is by renormalization group
effects$^{12)}$.  This has two immediate consequences:  first qualitatively the
SUSY
breaking scale is related to the electroweak mass scale (as appears to be the
case
experimentally from grand unification analysis).  More  quantitatively, the
renormalization
group equations allow one to relate the electroweak scale to the Gut scale.  As
a
consequence, the masses of the 32 new SUSY particles (listed in Table 1) can be
determined
in terms of only 4 additional Gut scale parameters, and the as yet unknown
t-quark mass
$m_t$.
\vfill\eject

\baselineskip=12pt
\itemitem{Table 1.} New particles predicted to exist in minimal SUSY models.
For squarks and
sleptons $i = 1,2,3$ is a generation index, $a$ is an $SU(3)_C$ index and
$\tilde{W}_i$,
$\tilde{Z}_i$ are labels so that $m_i < m_j$~for~$i < j$.
\medskip

\thicksize=0pt

{\begintable
\one & \one & \one & \one\cr
Name & Symbol & Type & Number\cr
squarks & $\tilde{q}_i = (\tilde{u}_{iL}, \tilde{d}_{iL});~\tilde{u}_{iR},
\tilde{d}_{iR}$
& $j = 0$, complex & 12\crnorule
\one & \one & \one & \one\crnorule
sleptons & $\tilde{l}_i = (\tilde{\nu}_{iL}, \tilde{e}_{iL});~\tilde{e}_{iR}$ &
$j = 0$,
complex & 9\crnorule
\one & \one & \one & \one\crnorule
gluino & $\lambda^a, a = 1 \cdots 8$ & $j = {1\over 2}$, Majorana & 1\crnorule
\one & \one & \one & \one\crnorule
Winos & $\tilde{W}_i;~i = 1,2$ & $j = {1\over 2}$, Dirac & 2\crnorule
(charginos) & \one & \one & \one\crnorule
\one & \one & \one & \one\crnorule
Zinos & $\tilde{Z}_i,~i = 1 \cdots 4$ & $j = {1\over 2}$, Majorana & 4\crnorule
(neutralinos) & \one & \one & \one\crnorule
\one & \one & \one & \one\crnorule
Higgs & $h^o, H^o$ & $j = 0$, real, CP even & 2\crnorule
\one & $A^o$ & $j = 0$, real, CP odd & 1\crnorule
\one & $H^{\pm}$ & $j = 0$, complex & 1\cr
& & &\endtable}

\thicksize=1.5pt

\medskip

\baselineskip=18pt
In principle then, if one knew the masses of 4 SUSY particles, one could
predict the
positions of the remaining 28 particles.  Of course, no SUSY particles have yet
been
discovered, and so in practice what one can do is determine various allowed
mass bands for
SUSY particles, or mass relations, between particles.  If the model possesses
proton decay,
existing (and future) bounds on the proton lifetime can considerably narrow
these bands.
Similarly, the cosmological constraint that the relic mass density of the
lightest
supersymmetric particle (which is stable is most models) not overclose the
universe, also
constrains the SUSY masses.  Thus it seems possible to test these models in the
relatively
near future.
\bigskip

\noindent
2.  CLASS OF MODELS
\medskip

We specify now the class of supergravity Gut models we will consider by
assuming the
following:

\item{(i)} There exists a hidden sector which is a gauge singlet with respect
to the
physical sector gauge group $G$ which breaks supersymmetry.  This can be done
by a super
Higgs mechanism$^{13)}$ or a gaugino condensate$^{14)}$.  The superpotential
$W$ is assumed
to decompose, e.g. for the super Higgs mechanism, as $W = W_{{\rm phys.}} (z_a)
+ W_{{\rm
hidden}} (z)$ where $\{z_a\}$ are the physical fields and $\{z\}$ the ($G$
singlet) hidden
sector fields.  The gauge hierarchy is maintained since the super Higgs fields
communicate
with the physical fields only gravitationally.

\item{(ii)} A Gut sector exists which breaks $G$ to the Standard Model group at
scale $Q =
M_G:  G \rightarrow SU(3)_C \times SU(2)_L \times U(1)_Y$.  An example of this
for the case
$G = SU(5)$ is given by the following Gut part of the superpotential$^{15)}$:

$$\eqalign{W_G=&\lambda_1 \bigg[{1\over 3} Tr \Sigma^3 + {1\over 2} M Tr
\Sigma^2\bigg]
+\cr
&\lambda_2 \bar{H}_{2X} (\Sigma^X_Y + 3M' \delta^X_Y) H_1\,^Y\cr}\eqno(2.1)$$

\item{}where $\Sigma^X_Y = 24$ of $SU(5)$, $\bar{H}_{2X}$~and~$H_1\,^Y$ are a
$\bar{5}$~and~$5$ of $SU(5)$.  Minimizing the effective potential one finds
diag
$<\Sigma^X_Y> = M (2,2,2,-3,-3)$, breaking $SU(5)$ with $M = O (M_G)$.  If $M'
= M + \mu_o/3
\lambda_2$, $\mu_o << M$, then the color triplet parts of $H_1$~and~$\bar{H}_2$
become
superheavy, and the $SU(2)$ doublets remain light and become the two Higgs
doublets of the
low energy theory.

\item{(iii)} After integrating out the superheavy fields, and eliminating the
super Higgs
fields, the only light particles remaining below the Gut scale are those of the
SUSY
Standard Model with one pair of Higgs doublets.

\item{(iv)} Any super Higgs field couplings that may appear in the Kahler
potential are
generation independent.

\noindent
Note that conditions (ii) and (iii) are just what is needed to obtain the grand
unification of the coupling constants discussed in Sec. 1, while (i) and (iv)
guarantees
the suppression of flavor changing neutral interactions.

A general model of this type can then be described at $M_G$ as follows$^{10)}$:
There is
an effective superpotential with quadratic and cubic terms $W = W^{(2)} +
W^{(3)}$ given
by

$$\eqalign{W = \mu_o H_1 H_2 + [&\lambda^{(u)}_{ij} q_i H_2 u^C_j +
\lambda^{(d)}_{ij} q_i H_1 d^C_j +\cr
&\lambda^{(e)}_{ij} l_i H_1 e^C_j],\cr}\eqno(2.2)$$

\noindent
an effective potential given by

$$\eqalign{V = \{\sum_a \mid {\partial W\over \partial z_a} \mid^2&+ V_D\} +
[m_o^2 \sum_a
z_a z_a^{\ast} +\cr
&(A_o W^{(3)} + B_o W^{(2)} + h.c.)]\cr}\eqno(2.3)$$

\noindent
and a universal gaugino mass term ${\cal L}^{\lambda}_{{\rm mass}} = -m_{1/2}
\bar{\lambda}^{\alpha} \lambda^{\alpha}$.  In Eq. (2.2), $q_i, l_i, H_1, H_2$
are
$SU(2)_L$ doublets, $u_i^C$, $d_i^C$, $e_i^C$ are conjugate singlets, $V_D$ is
the usual
$D$ term, and $\lambda^{(u)}$, $\lambda^{(d)}$, $\lambda^{(e)}$ are the usual
Yukawa
coupling constants.  In Eq. (2.3), $m_o^2 > 0$ is a universal mass term for all
scalar
fields.  Thus aside from the Yukawa coupling constants of the Standard Model,
the theory
depends on the following Gut scale parameters:

$$m_{1/2}, m_o, A_o, B_o;~\mu_o;~\alpha_G, M_G\eqno(2.4)$$

\noindent
The first four constants are the ``soft-breaking'' parameters that characterize
supersymmetry breaking, and $\mu_o$ is the $H_1 - H_2$ mixing parameter.
\bigskip

\noindent
3.  ELECTROWEAK BREAKING
\medskip

We briefly summarize next how the supergravity models give rise naturally to
electroweak
breaking.  At $Q = M_G$, we saw in Sec. 2 that the spontaneous breaking of
supersymmetry
gave all scalar fields a universal mass $m_o$ where $m_o^2 > 0$.  Using the
renormalization group equations (RGE), each particle's mass changes due to
radiative
corrections as one goes to lower values of $Q$.  The squark, slepton and $H_1$
(mass)$^2$
increase, but due to the t-quark Yukawa couplings the $H_2$ (mass)$^2$ is
driven negative
at the electroweak scale, as shown schematically in Fig. 1.  To see this
\vskip 2.75truein

\baselineskip=12pt
\itemitem{Fig. 1.} Running masses in supersymmetric models as a function of the
mass scale
$Q$.
\bigskip

\baselineskip=18pt
\noindent
in more detail, consider the part of the effective potential of Eq. (2.3)
involving the
Higgs fields:

$$\eqalign{V_H=&m_1 (t)^2 \mid H_1 \mid^2 + m_2 (t)^2 \mid H_2 \mid^2 - m_3^2
(t) (H_1 H_2
+ h.c.) +\cr
&{1\over 8} [g_2^2 (t) + g_Y^2 (t)] [\mid H_1 \mid^2 - \mid H_2 \mid^2]^2 +
\Delta
V_1\cr}\eqno(3.1)$$

\noindent
where $t = ln [M_G^2/Q^2]$ is the running parameter, $m_i^2 (t) = m_{H_i}^2 (t)
+ \mu^2 (t),
i = 1,2, m_3^2 (t) = - B (t) \mu (t)$~and~$\Delta V_1$ is the one loop
correction.  At $Q =
M_G~(t = 0)$, the running masses obey the boundary conditions $m_i^2 (0) =
m_o^2 +
\mu_o^2$~and~$m_3^2 (0) = - B_o \mu_o$.  The RGE determine these parameters at
all other
$t$.  One may minimize $V_H$ with respect to the two VEVs $\sigma_{1,2} =
<H_{1,2}>$ to
obtain

$${1\over 2} M_Z^2 = {\mu_1^2 - \mu_2^2 \tan^2 \beta\over \tan^2 \beta -
1}~;~\sin 2\beta =
{2 m_3^2\over \mu_1^2 + \mu_2^2}\eqno(3.2)$$

\noindent
where $\mu_i^2 = m_i^2 + \Sigma_i, \tan \beta = v_2/v_1$~and~$\Sigma_i$ are the
loop
corrections:  $\Sigma_i = \Sigma_a (-1)^{2j_a}$\break\hfil $n_a [M_a (v_i)]^2
ln
[M_a^2/\sqrt{e} Q^2] {\partial M_a^2\over \partial v_i}$.  ($M_a$ is the mass
of particle
$a$, $j_a$ is its spin and $n_a$ is the number of helicity states.).  In
practice, Eqs.
(3.2) are insensitive to the value of $Q^2$ in the electroweak scale$^{16)}$ so
one may
conveniently set $Q = M_Z$.  Also, the loop corrections are generally
small$^{16)}$.

The RGE allow one to evaluate $\mu_i^2 (t)$~and~$m_3^2 (t)$ in terms of the Gut
parameters.  From the boundary conditions above, one may use Eqs. (3.2) to
elimate
$\mu_o^2$ in terms of $M_Z$ and replace $B_o$ by $\tan \beta$.  One is left
with the
parameters

$$m_o, m_{1/2}, A_o, \tan \beta;~\alpha_G, M_G\eqno(3.3)$$

\noindent
The sign of $\mu_o$ is not determined and so there are two branches:  $\mu_o >
0$~and~$\mu_o < 0$.  Since $\alpha_G$~and~$M_G$ have essentially been
``measured'' by LEP
in the grand unification analysis of Sec. 1, the theory depends on 4 + 1
constants:  $m_o,
m_{1/2}, A_o, \tan \beta$ and the as yet undetermined $m_t$.  For a fixed set
of these
parameters, one can calculate the masses of all the SUSY particles.  An example
of this is
given in Fig. 2
\vfill\eject

\null
\vskip 5.5truein

\baselineskip=12pt
\itemitem{Fig. 2.} The SUSY mass spectrum for $m_o = 600$ GeV, $m_{1/2} = 53$
GeV, $A_o = 0$,
$\tan \beta = 1.73$, $\mu < 0$~and~$m_t = 150$ GeV.
\bigskip

\baselineskip=18pt
\noindent
Note the mass splitting in the third generation of squarks, in the Winos and
Zinos, and in
the neutralinos.

One can vary all the parameters, and in this way get allowed bands of SUSY
masses.  In the
following, we will also impose a theoretical constraint that there will be no
excessive
``fine tuning'' of parameters, which we will take as requiring $m_o,
m_{\tilde{g}} < 1$
TeV.  This also implies that squarks and gluinos lie below 1 TeV, which is also
probably
the upper limit for detecting these particles at the SSC or LHC.
\bigskip

\noindent
4.  PROTON DECAY
\medskip

We consider here models with ``$SU(5)$-type'' proton decay.  These are models
which obey
the following conditions:  (i) The Gut group $G$ contains an $SU(5)$ subgroup
[or is
$SU(5)$].  (ii) The matter that remains light after $G$ breaks to $SU(3) \times
SU(2)
\times U(1)$ at $M_G$ is embedded in the usual way in the $10 + \bar{5}$
representations of
the $SU(5)$ subgroup.  (iii) After $G$ breaks, there are only two light Higgs
doublets
which interact with matter, and these are embedded in the $5 + \bar{5}$ of the
$SU(5)$
subgroup.  The corresponding Higgs color triplets are assumed to become
superheavy from a
$M_{H_3} H_3 \bar{H}_3$ term arising after the breaking of $G$.  (iv) There is
no discrete
symmetry or condition that forbids the proton decay amplitude.

Under the above conditions (which can arise in a number of models, e.g. $G =
SU(5), O(10),
E_6$ etc.)  There is a characteristic SUSY proton decay, $p \rightarrow
\bar{\nu} + K^+$, due
to the exchange of the superheavy Higgsino color triplet with a model
independent decay
amplitude$^{17,18)}$.  An example of this decay process is given in Fig. 3.
Proton
\vskip 2.25truein

\baselineskip=12pt
\itemitem{Fig. 3.} One of the diagrams contributing to the $p \rightarrow
\bar{\nu}_{\mu} +
K^+$ decay.  The Wino ($\tilde{W}$) converts the quarks into squarks, and the
baryon
violating interactions occur at the $\tilde{H}_3$ vertex.
\bigskip

\baselineskip=18pt
\noindent
decay is a characteristic feature of supergravity grand unification models, and
one must do
special things to avoid it.  Thus the flipped $SU(5)$ model suppresses proton
decay by
violating condition (2) above$^{19)}$.  Models that invoke discrete symmetries
to prevent
$p$-decay from arising  generally have more than one pair of light Higgs
doublets
and sometimes relatively light Higgs color triplets$^{20)}$.  While proton
decay would be
suppressed, one would expect such models to be in disagreement with the LEP
grand
unification data, which  requires only one pair of light Higgs doublets$^{8)}$.

The current experimental bound on the $p \rightarrow \bar{\nu} K^+$ mode is,
from
Kamiokande$^{21)}$, $\tau (p \rightarrow \bar{\nu} K^+) > 1 \times 10^{32}$ yr
(90\% CL).
However, future experiments can greatly improve on this limit and are expected
to be
sensitive up to $\simeq 2 \times 10^{33}$ yr for Super
Kamiokande$^{22)}$~and~$\simeq 5
\times 10^{33}$ yr for ICARUS$^{23)}$.

The total decay rate is $\Gamma (p \rightarrow \bar{\nu} K) = \Sigma_i \Gamma
(p \rightarrow
\bar{\nu}_i K), i = e, \mu, \tau$.  The CKM matrix elements appear at the
vertices of the
loop integral of Fig. 3 and so all three generations can circle in the loop.
Thus for a
superheavy $\tilde{H}_3$, one may write$^{18)}$

$$\Gamma (p \rightarrow \bar{\nu} K) = {\rm Const} (\beta_p/M_{H_3})^2
\sum_{a,i} \mid
B_{ia} \mid^2\eqno(4.1)$$

\noindent
where $B_{ia}$ is the loop amplitude of the $\bar{\nu}_i K$ mode when
generation $a$ squarks
enter in the loop.  (Actually, the first generation, $i = 1$~and~$a = 1$, give
negligible
contributions.)  The quantity $\beta_p$ is

$$\beta_p U_L^{\gamma} = \varepsilon_{abc} \epsilon_{\alpha\beta} <o \mid
d_{aL}^{\alpha}
u_{bL}^{\beta} u_{cL}^{\gamma} \mid p>\eqno(4.2)$$

\noindent
where $U_L^{\gamma}$ is the proton wave function.  Lattice gauge calculations
give$^{24)}$
$\beta_p = (5.6 \pm 0.8) \times 10^{-3}$ GeV$^{-1}$.  The general expression
for the loop
amplitudes $B_{ia}$ are complicated functions given in Ref. (18).  They clearly
depend on
the SUSY particle ($\tilde{q}, \tilde{W}, \tilde{l}$) masses, and so an upper
bound on
$\Gamma (p \rightarrow \bar{\nu} K)$ will produce bounds on the SUSY masses.
However,
$M_{H_3}$ also enters in $\Gamma$, and one also needs information concerning
this
quantity.  In general one expects $M_{H_3} = O (M_G)$, and so to quantify the
relation we
first return to reconsider the grand unification of the coupling constants
$\alpha_1,
\alpha_2, \alpha_3$.
\bigskip

\noindent
5.  UNIFICATION OF COUPLING CONSTANTS
\medskip

The analysis of the unification of $\alpha_1, \alpha_2$~and~$\alpha_3$ is
complicated by
the existence of two sets of thresholds that exist as one proceeds from $M_Z$
to $M_G$
[using the renormalization group equations (RGE)].  There are first the low
energy
thresholds due to the spectrum of SUSY particles at masses $\sim 100$ GeV - 1
TeV, and
second there are the superheavy Gut particles as masses $\sim M_G$ that account
for the
breaking of the Gut group $G$ to $SU(3) \times SU(2) \times U(1)$.  If, as a
zero'th
approximation, one sets all SUSY particles to a common, ``average'' mass $M_S$,
and all Gut
particles to $M_G$, then a fit to the data $\alpha_1 (M_Z), \alpha_2 (M_Z),
\alpha_3 (M_Z)$
gives

$$M_G = 10^{16.19 \pm 0.34} {\rm GeV};~M_S = 10^{2.37 \pm 1.0} GeV\eqno(5.1)$$

\noindent
and $\alpha_G^{-1} = 25.7 \pm 1.7$, the errors being due to those in $\alpha_3$
which we
will take here as$^{24)}$

$$\alpha_3 (M_Z) = 0.118 \pm 0.007\eqno(5.2)$$

It is possible, when the Gut mass spectrum is taken into account that $M_{H_3}$
will exceed
the above value for $M_G$.  To get some idea on how big $M_{H_3}$ could be,
consider the
$SU(5)$ model for the Gut sector of Eq. (2.1).  One finds, after the breaking
of $SU(5)$
that $M_{H_3} = 5 \lambda_2 M$, the octet and singlet component of the
\underbar{24}
have masses $M_{\Sigma}^{(8,3)} = 5 \lambda_1 M/2$, $M_{\Sigma}^o = \lambda_1
M/2$, and the
massive vector bosons have mass $M_V = 5 gM (\alpha_G = g^2/4\pi)$.  To stay
within the
perturbative domain we restrict $\lambda_{1,2} \leq 2$ (i.e.
$\alpha_{\lambda_{1,2}} =
\lambda_{1,2}^2/4\pi \l 1/3$).  We also limit $\lambda_{1,2} > 0.01$ (i.e.
$\alpha_{\lambda_{1,2}} > 8 \times 10^{-6}$).  One may now carry out the grand
unification
analysis including the Gut thresholds.  The result for the allowed region is
given in Fig.
4.  We note first that grand unification implies an {\it upper} bound on
$\alpha_3$ of
\vskip 3.0truein

\baselineskip=12pt
\itemitem{Fig. 4.} Relation between Higgs triplet mass $M_{H_3}$~and~$\alpha_3
(M_Z)$
required by grand unification for the Gut model of Eq. (2.1).$^{25)}$  The
quadrilateral
region enclosed by the solid lines is the allowed region consistent with grand
unification
for 30 GeV $< M_S <$ 1 TeV, $\lambda_1 > 0.01, \lambda_2 < 2$.
\bigskip

\baselineskip=18pt
\noindent
$\alpha_3 (M_Z) \l 0.135$ (which is reduced for a larger value of $\lambda_1$),
while the $1
- \sigma$ bound of Eq. (5.2), $\alpha_3 (M_Z) = 0.125$, corresponds to $M_{H_3}
\simeq 2
\times 10^{17}$ GeV or $M_{H_3} \simeq 10~M_G$.  In the following, we will
assume

$$3 < M_{H_3}/M_G < 10\eqno(5.3)$$

\noindent
as a reasonable range for $M_{H_3}$.
\bigskip

\noindent
6.  SUSY MASS RELATIONS
\medskip

We now examine the SUSY mass spectrum obtained by letting the parameters of the
theory,
$m_o, m_{1/2}, A_o, \tan\beta$~and~$m_t$ range over the entire parameter space
subject
only to the following constraints:  (i) the SUSY masses and $m_t$ do not
violate current
experimental bounds; (ii) Radiative breaking of $SU(2) \times U(1)$ occurs
(i.e. solutions
of Eqs. (3.2) exist); (iii) Experimental bounds on proton decay are obeyed;
(iv) No
excessive fine tuning occurs i.e. $m_o, m_{\tilde{g}} < 1$ TeV, and $M_{H_3}$
is
constrained by Eq. (5.3).  We summarize now the consequences of the model under
these
conditions.

(1) We examine first the smallest value of $M_{H_3}$, i.e. $M_{H_3}/M_G = 3$,
where
proton decay is most constraining.  The parameter space is limited but still
sizable.  One
finds$^{1)}$

$$\eqalign{m_o \r 500~&{\rm GeV};~m_{\tilde{g}} \l 450~{\rm GeV};~- 1.5 \l
A_t/m_o
\l 1.5\cr
&1.1 \l \tan \beta \l 5\cr}\eqno(6.1)$$

\noindent
This implies that squarks (except perhaps $\tilde{t}_1$, the light t-squark)
and probably
gluinos will require the SSC and LHC to be seen.  In addition, one finds the
bounds $m_t
< 180$ GeV and $m_h < 110$ GeV.  Further$^{2)}$, for $m_t < 140$ GeV, one finds
that
$m_{\tilde{W}_1} < 100$ GeV whenever $m_h < 95$ GeV.  Since these are the
respective
bounds for observing the $\tilde{W}_1$~and~$h$ particles at LEP2, one has that
if $m_t <
140$ GeV, LEP2 will see either the $\tilde{W}_1$ or the $h$ (and possibly
both).

(2) As $M_{H_3}/M_G$ increases, the lower bound on $m_o$ decreases and the
upper bound on
$m_{\tilde{g}}$ increases.  Thus for $M_{H_3}/M_G \r 7$, $m_{\tilde{g}}$ can
reach the
maximum allowed value of 1 TeV.  One will still generally expect to need the
SSC or LHC
to detect squarks and the gluino.  (The other bounds of Eq. (6.1) also widen,
through not
greatly.)

(3) Over most of the allowed parameter space, for the whole range of $M_{H_3}$
of Eq.
(5.3), a remarkable set of scaling laws hold for the light charginos and
neutralinos$^{1-3)}$:
$$\eqalignno{2 m_{\tilde{Z}_1}&\cong m_{\tilde{W}_1} \cong
m_{\tilde{Z}_2}&(6.2a)\cr
m_{\tilde{W}_1}&\simeq {1\over 4} m_{\tilde{g}} (\mu > 0);~m_{\tilde{W}_1}
\simeq {1\over
3} m_{\tilde{g}} (\mu < 0)&(6.2b)\cr}$$

\noindent
(Eqs. (6.2a) often hold to within a few percent and Eqs. (6.2b) to within
25\%.)  In
addition, the other chargino and neutralinos are nearly degenerate and much
heavier than
the $\tilde{Z}_1$.  Similarly, the other Higgs bosons are generally very heavy
and nearly
degenerate:
$$\eqalignno{m_{\tilde{W}_2}&\cong m_{\tilde{Z}_3} \cong m_{\tilde{Z}_4} >>
m_{\tilde{Z}_1}&(6.3a)\cr
m_H&\cong m_A \cong m_{H^{\pm}} >> m_h&(6.3b)\cr}$$

\noindent
The reason for Eqs. (6.2) and (6.3a), is that generally one finds that the
proton decay
constraint requires $\mu^2 >> M_Z^2, \tilde{m}_2^2$ (where $\tilde{m}_2$ is the
$SU(2)$
gaugino mass) while Eq. (6.3b) is a consequence of the largeness of $m_o$.
\bigskip

\noindent
7.  FUTURE EXPERIMENTS
\medskip

One can combine the expectations from future experiments to obtain fairly
stringent tests
for these models.  Thus Super Kamiokande expects to reach a sensitivity
of$^{22)}$ $\approx
2 \times 10^{33}$ yr for the $p \rightarrow \bar{\nu} K^+$ mode, while ICARUS
expects to
reach to$^{23)}$ $\approx 5 \times 10^{33}$ yr.  Figs. 5 show the maximum value
of $\tau (p
\rightarrow \bar{\nu} K^+)$ as a function of $m_o$ as all other parameters are
varied over
the entire allowed parameter space.
\vskip 3.25truein

\baselineskip=12pt
\itemitem{Fig. 5a.} The maximum value of $\tau (p \rightarrow \bar{\nu} K^+)$
vs $m_o$ for
$m_t = 125$ GeV, $\mu < 0$.  The maximum is calculated by varying all
parameters except
$m_o$ over the entire allowed parameter space.  The results are plotted for
$M_{H_3}/M_G =$
3, 6 and 10.  The lower horizontal line is the upperbound for Super Kamiokande,
and the
higher line is for ICARUS.
\bigskip

\baselineskip=18pt
\noindent
One sees that the entire domain for $m_o \l 1000$ GeV is excluded by ICARUS for
\break\hfil $M_{H_3}/M_G \l 6$ if proton decay is not observed.  (The same
result holds for
Super~~~\break\hfil Kamiokande with $m_o \l 800$ GeV.)
\vfill\eject

\null
\vskip 3.0truein

\baselineskip=12pt
\itemitem{Fig. 5b.} The same as Fig. 5a for $m_t = 150$ GeV, $\mu < 0$.
\vskip 3.25truein

\itemitem{Fig. 5c.} The same as Fig. 5a for $m_t = 170$ GeV, $\mu < 0$.
\bigskip

\baselineskip=18pt
Fig. 6 plots the maximum value of $\tau (p \rightarrow \bar{\nu} K^+)$ for $m_o
= 400$ GeV,
800 GeV and 1200 GeV as a function of $m_t$.  This lifetime peaks at $m_t
\simeq 145$ GeV.
The reason for this arises from
\vfill\eject

\null
\vskip 3.0truein

\baselineskip=12pt
\itemitem{Fig. 6.} Maximum value of $\tau (p \rightarrow \bar{\nu} K^+)$ vs
$m_t$ for
$M_{H_3}/M_G = 6$~and~$\mu < 0$.  The solid line is for $m_o = 400$ GeV, the
dashed line
for $m_o = 800$ GeV and the dot-dashed line for $m_o = 1200$ GeV.  The lower
horizontal
line is the upper bound that Super Kamiokande can detect, and the higher
horizontal line is
the upper bound for ICARUS.
\bigskip

\baselineskip=18pt
\noindent
two competing phenomena:  As $m_t$ increases, the off-diagonal terms of the
t-squark mass
matrix, $m_t (A_t m_o + \mu~{\rm ctn}~\beta)$, increases, reducing the
$\tilde{t}_i$ mass
and allowing more destructive interference between the third and second
generation
contributions to the loop of Fig. 3.  However, for large $m_t$, the allowed
parameter space
shrinks (e.g. $A_t$ approaches zero) reducing the off-diagonal terms again.
Note also that
Fig. 6 shows that Super Kamiokande is accessible to the parameter space when
$m_o \l 800$
GeV and $M_{H_3}/M_G < 6$.

Fig. 7 shows the maximum value of $\tau (p \rightarrow \bar{\nu} K^+)$ as a
function of
$m_o$ for $m_t = 150$
\vfill\eject

\null
\vskip 3.0truein

\baselineskip=12pt
\itemitem{Fig. 7.} Maximum value of $\tau (p \rightarrow \bar{\nu} K^+)$ vs
$m_o$ for
$M_{H_3}/M_G = 3$ (solid line),\break\hfil $M_{H_3}/M_G = 6$ (dashed line),
$M_{H_3}/M_G =
10$ (dot-dashed line) when $m_t = 150$ GeV, $\mu > 0$~and~$m_{\tilde{W}_1} >
100$ GeV.  The
horizontal lines are as in Figs. 5,6.
\bigskip

\baselineskip=18pt
\noindent
GeV, $\mu > 0$ (which is near the maximum of the Fig. 6 curves) subject to
the constraint that $m_{\tilde{W}_1}$ be greater than 100 GeV (and hence not be
accessible
to LEP 200).  The lifetime increases with increasing Wino mass, and as can be
seen in Fig.
7, it implies that even for $M_{H_3}/M_G < 10$, proton decay should be
accessible to ICARUS
for $m_o \l 1250$ GeV (and accessible to Super Kamiokande for $m_o \l 950$ GeV)
if the
$\tilde{W}_1$ is not seen at LEP 200.  Thus one or the other of these signals
for this
class of models should be accessible experimentally.
\bigskip

\noindent
8.  SUMMARY AND CONCLUSIONS
\medskip

Supergravity grand unification models depend on relatively few additional
parameters, and
consequently have a good amount of predictive power.  For models possessing
$SU(5)$-type
proton decay, the new round of planned proton decay experiments combined with
LEP 200 can
give strong tests of these Gut models.  One finds

\item{(i)} For Gut models with $M_{H_3}/M_G < 6$ the decay mode $p \rightarrow
\bar{\nu} K^+$
should be seen at ICARUS for the entire range $m_o, m_{\tilde{g}} < 1$ TeV (and
be seen at
Super Kamiokande for $m_o < 800$ GeV).

\item{(ii)} For $M_{H_3}/M_G < 10$~and~$m_o < 1250$ GeV, $m_{\tilde{g}} < 1$
TeV, either
the mode $p \rightarrow \bar{\nu} K^+$ would be seen at ICARUS or the
$\tilde{W}_1$ has mass
$m_{\tilde{W}_1} < 100$ GeV and hence should be observable at LEP 200.
(Similarly for
Super Kamiokande for $m_o < 950$ GeV).

\item{(iii)} For $M_{H_3}/M_G < 10$~and~$m_o, m_{\tilde{g}} < 1$ TeV one finds
that if
$\tau (p \rightarrow \bar{\nu} K^+) > 1.5 \times 10^{33}$ yr, then either $m_h
< 95$ GeV or
$m_{\tilde{W}_1} < 100$ GeV.  Thus either the $h$ or the $\tilde{W}_1$ (and
possibly both)
would be observable at LEP 200.  (Note that the condition $\tau > 1.5 \times
10^{33}$ GeV
could be tested at both Super Kamiokande and ICARUS.)

In addition to the above, over most of the allowed parameter space, we expect
the gaugino
scaling relations, Eqs. (6.2), and the degeneracy relations, Eqs. (6.3), to
hold.  While the
SSC or LHC are probably needed to see the gluino and squarks, Eqs. (6.2) allow
for the
possibility of detection of light gauginos and the light $h$ Higgs at the
Tevatron and LEP
200.

Models of the type we have been considering possess R parity invariance, and as
a
consequence, the lightest supersymmetric particle (LSP) is totally stable.  The
proton
decay constraint implies that the LSP be the $\tilde{Z}_1$.  Cosmological
constraints then
require that the relic density of the LSP be sufficiently small that it not
over close the
universe.  The dominant annihilation processes in the early universe occur
mainly via the
s-channel $h$~and~$Z$ poles.  Recent detailed calculations show$^{26)}$ that
the relic
density constraint can be viewed as a bound on the allowed gluino mass region.
Allowed
gluino
\vskip 2.75truein

\baselineskip=12pt
\item{Fig. 8.} Region in $m_{\tilde{g}} - A_t$ space allowed by the combined
proton decay
and cosmological constraints for $m_t = 125$ GeV, $m_o = 600$ GeV, $\tan \beta
= 1.73, \mu
> 0$~and~$M_{H_3}/M_G = 6$.  The lower band is due to the Higgs pole, and the
upper band is
due to the $Z$ pole.
\bigskip

\baselineskip=18pt
\noindent
mass bands of $\approx 40$ GeV arise from the $h$ pole and $\approx 20$ GeV
from the $Z$
pole.  Sometimes these two regions merge giving a broad band of allowed values
of
$m_{\tilde{g}}$.  Further, one finds $m_t \l 165$ GeV, $m_h < 105$ GeV,
$m_{\tilde{W}_1} <
100$ GeV and $m_{\tilde{Z}_1} < 50$ GeV for $M_{H_3}/M_G < 6$.  Thus while the
cosmological
constraint does indeed further limit the parameter space of supergravity grand
unified
models, there still remains a sizable allowed region.  It should be stressed
that should
even one of the above considered signals be experimentally observed (e.g. a
light Higgs, or
proton decay) one will be able to use this new data to give even more precise
predictions
that could test the validity of these models.
\bigskip

\noindent
ACKNOWLEDGEMENTS:  This research was supported in part by NSF Grant Nos.
PHY-916593 and
PHY-917809.
\bigskip

\noindent
REFERENCES
\bigskip

\item{1.} R. Arnowitt and P. Nath, Phys. Rev. Lett. {\bf 69}, 725 (1992).

\item{2.} P. Nath and R. Arnowitt, Phys. Lett. {\bf B289}, 368 (1992).

\item{3.} S. Kelley, J. Lopez, D. V. Nanopoulos, H. Pois and K. Yuan, Phys.
Lett. {\bf
B273}, 423 (1991).

\item{4.} P. Nath and R. Arnowitt, Phys. Lett. {\bf B287}, 3282 (1992).

\item{5.} G. G. Ross and R. G. Roberts, Nucl. Phys. {\bf B377}, 971 (1992).

\item{6.} M. Drees and M. M. Nojiri, Nucl. Phys. {\bf B369}, 54 (1992).

\item{7.} K. Inoue, M. Kawasaki, M. Yamaguchi and T. Yanagida, Phys. Rev. {\bf
D45}, 387
(1992).

\item{8.} P. Langacker, Proc. PASCOS 90 Symposium, p. 231, Eds. P. Nath and S.
Reucroft
(World Scientific, Singapore, 1990), D. Langacker and M. Luo, Phys. Rev. {\bf
D44}, 817
(1991); J. Ellis, S. Kelley and D. V. Nanopoulos, Phys. Lett. {\bf B249}, 447
(1991); {\bf
B260}, 131 (1991); F. Anselmo, L. Cifarelli, A. Peterman and A. Zichichi, Nuovo
Cim {\bf
104A}, 1817 (1991); {\bf 115A}, 581 (1992).

\item{9.} R. Barbieri, S. Ferrara and C. A. Savoy, Phys. Lett. {\bf B119}, 343
(1982).

\item{10.} L. Hall, J. Lykken and S. Weinberg, Phys. Rev. {\bf D27}, 2359
(1983); P. Nath,
R. Arnowitt and A. H. Chamseddine, Nucl. Phys. {\bf B227}, 121 (1983); S. Soni
and A.
Weldon, Phys. Lett. {\bf B126}, 215 (1983).

\item{11.} A. H. Chamseddine, R. Arnowitt and P. Nath, Phys. Rev. Lett. {\bf
49}, 970
(1982).

\item{12.} K. Inoue et al., Prog. Theor. Phys. {\bf 68}, 927 (1982); L.
Iba\~nez and G. G.
Ross, Phys. Lett. {\bf B110}, 227 (1982); L. Avarez-Gaum\'e, J. Polchinski and
M. B. Wise,
Nucl. Phys. {\bf B250}, 495 (1983); J. Ellis, J. Hagelin, D. V. Nanopoulos and
K. Tamvakis,
Phys. Lett. {\bf 125B}, 275 (1983); L. E. Iba\~nez and C. Lopez, Phys. Lett.
{\bf B128}, 54
(1983); Nucl. Phys. {\bf B233}, 545 (1984); L. Iba\~nez, C. Lopez, and C.
Mu\~nos, Nucl.
Phys. {\bf B256}, 218 (1985).

\item{13.} J. Polonyi, Univ. Budapest Report KFKI-1977-93 (1977).

\item{14.} P. Nilles, Phys. Lett. {\bf B115}, 193 (1981).

\item{15.} E. Witten, Nucl. Phys. {\bf B177}, 477 (1981); {\bf B185}, 513
(1981); S.
Dimopoulos and H. Georgi, Nucl. Phys. {\bf B193}, 150 (1981); N. Sakai, Zeit.
f. Phys.
{\bf C11}, 153 (1981).

\item{16.} G. Gamberini, G. Ridolfi and F. Zwirner, Nucl. Phys. {\bf B331}, 331
(1990); R.
Arnowitt and P. Nath, Phys. Rev. {\bf D46}, 3981 (1992).

\item{17.} S. Weinberg, Phys. Rev. {\bf D26}, 287 (1982); N. Sakai and T.
Yanagida, Nucl.
Phys. {\bf B197}, 533 (1982); S. Dimopoulos, S. Raby and F. Wilczek, Phys.
Lett. {\bf
112B}, 133 (1982); J. Ellis, D. V. Nanopoulos and S. Rudaz, Nucl. Phys. {\bf
B202}, 43
(1982); S. Chadha and M. Daniels, Nucl. Phys. {\bf B229}, 105 (1983); B. A.
Campbell, J.
Ellis and D. V. Nanopoulos, Phys. Lett. {\bf 141B}, 229 (1984).

\item{18.} R. Arnowitt, A. H. Chamseddine and P. Nath, Phys. Lett. {\bf B156},
215 (1985);
P. Nath, R. Arnowitt and A. H. Chamseddine, Phys. Rev. {\bf D32}, 2348 (1985).

\item{19.} I. Antoniadis, J. Ellis, J. Hagelin and D. V. Nanopoulos, Phys.
Lett. {\bf
B208}, 209 (1988).

\item{20.} G. D. Coughlan et al., Phys. Lett. {\bf B160}, 249 (1985); A.
Masiero et al.,
Phys. Lett. {\bf B115}, 380 (1982).

\item{21.} Particle Data Group, Phys. Rev. {\bf D45}, Part 2 (1992).

\item{22.} Y. Totsuka, Proc. XXIV Conf. on High Energy Physics, Munich, 1988,
Eds. R.
Kotthaus and J. H. Kuhn (Springer Verlag, Berlin, Heidelberg, 1989).

\item{23.} ICARUS Detector Group, Int. Symposium on Neutrino Astrophysics,
Takayama, 1992.

\item{24.} M. B. Gavela et al., Nucl. Phys. {\bf B312}, 269 (1989).

\item{25.} D. Ring and S. Urano.

\item{26.} R. Arnowitt and P. Nath, Phys. Lett. {\bf B299}, 58 (1993) and
Erratum; P. Nath
and R. Arnowitt, NUB-TH-3056/92-CTP-TAMU-66/92-TIFR/TH/92-69-SSCL-Preprint-225
(Revised).
\vfill\eject

Fig. 1\qquad $m_o\,^2$ \qquad $M_Z$ \qquad $M_G$
\bigskip

Fig. 3
$p$   $\Bigg\{$   $u$    $\tilde{d}$     $\bar{\nu}_{\mu}$

$\tilde{W}$    $\hat{u}$    $\tilde{H}_3$

$d$    $\bar{s}$    $\Bigg\}$    $K^+$

$u$   $u$   $\Bigg\}$
\bigskip

Fig. 5a
Log$_{10}$ [$\tau (p \rightarrow \bar{\nu} K)_{{\rm max}}$]
\bigskip

$m_o$ (GeV)
\bigskip

Fig. 5b
Log$_{10}$ [$\tau (p \rightarrow \bar{\nu} K)_{{\rm max}}$]
\bigskip

$m_o$ (GeV)
\bigskip

Fig. 5c
Log$_{10}$ [$\tau (p \rightarrow \bar{\nu} K)_{{\rm max}}$]
\bigskip

$m_o$ (GeV)
\bigskip

Fig. 6
Log$_{10}$ [$\tau (p \rightarrow \bar{\nu} K)_{{\rm max}}$]
\bigskip

$m_t$ (GeV)
\bigskip

Fig. 7
Log$_{10}$ [$\tau (p \rightarrow \bar{\nu} K)_{{\rm max}}$]
\bigskip

$m_o$ (GeV)
\vfill\eject\bye